\documentclass[twocolumn,aps,prd,groupedaddress,nofootinbib,showpacs]{revtex4}%
\usepackage{graphicx}
\usepackage{amsmath}
\usepackage{bm}

\usepackage{relsize}
\RequirePackage{xspace}
\usepackage{dcolumn}
\usepackage{bm}

\usepackage{relsize}
\RequirePackage{xspace}
\usepackage{dcolumn}
\usepackage{bm}

\def\jpsi{{J/\psi}}
\def\psip{{\psi^\prime}}

\def\sa{{\bigl.^1\hspace{-1mm}S^{[8]}_0}}
\def\sb{{\bigl.^3\hspace{-1mm}S^{[8]}_1}}
\def\pj{{\bigl.^3\hspace{-1mm}P^{[8]}_J}}
\def\mo{\mathcal{O}}

\def\mopa{{\langle\mathcal{O}^\jpsi(\bigl.^1\hspace{-1mm}S_0^{[8]})\rangle}}
\def\mopb{{\langle\mathcal{O}^\jpsi(\bigl.^3\hspace{-1mm}S_1^{[8]})\rangle}}
\def\mopc{{\langle\mathcal{O}^\jpsi(\bigl.^3\hspace{-1mm}P_0^{[8]})\rangle}}

\def\dssa{{\rm d}\hat{\s}[\sa]}
\def\dssb{{\rm d}\hat{\s}[\sb]}
\def\dspj{{\rm d}\hat{\s}[\pj]}

\def\Majpsi{M_{0,r_0}^{\jpsi}}
\def\Mbjpsi{M_{1,r_1}^{\jpsi}}

\def\MoH{{\langle\mathcal{O}^H\rangle}}
\def\MaH{M_{0,r_0}^{H}}
\def\MbH{M_{1,r_1}^{H}}
\def\ptcut{p_T^{\text{cut}}}
\def\be{\begin{equation}}
\def\ee{\end{equation}}
\def\bea{\begin{eqnarray}}
\def\eea{\end{eqnarray}}
\def\NO{\nonumber}

\def\gev{\mathrm{~GeV}}
\def\tev{\mathrm{~TeV}}

\def\dfrac{\displaystyle\frac}

\def\vv{\overrightarrow{\textbf{v}}}
\def\EO{\Lambda}

\def\a{\alpha}

\def\s{\sigma}

\def\muL{\mu_{\Lambda}}

\def\scut{s_{ij}^{\text{min}}}
\begin{document}


\title{\mbox{}\\[10pt] A complete NLO calculation of the $\bm{J/\psi}$ and $\bm{\psip}$ production at hadron colliders\footnote{An expanded version based on Ref.\cite{talk}.}}

\author{Yan-Qing Ma}
\author{Kai Wang}
\affiliation{Department of Physics and
State Key Laboratory of Nuclear Physics and Technology, Peking
University, Beijing 100871, China}

\author{Kuang-Ta Chao}
\affiliation{Department of Physics
and State Key Laboratory of Nuclear
Physics and Technology, and Center for High Energy Physics,
Peking  University, Beijing 100871, China}

\date{\today}

\begin{abstract}
A complete next-to-leading order (NLO) calculation in $\alpha_s$ for
the $\jpsi$ and $\psip$ prompt production at the Tevatron, LHC, and
RHIC in nonrelativistic QCD is presented. We argue that the
next-to-next-to-leading order (NNLO) color-singlet contribution may
not be so important as to resolve the large discrepancy between
theory and experiment in $\jpsi$ large $p_T$ production cross
sections. Therefore, a complete NLO calculation, including both
color-singlet and color-octet contribution, is necessary and
essential to give a good description for $\jpsi$ and $\psip$
production. We also study the methods to fit the long-distance
matrix elements using either two linear combined matrix elements or
three matrix elements, and find these two methods can give
consistent results. Compared with the measurements at the LHC and
RHIC for prompt $\jpsi$ and $\psip$ production, our predictions are
found to agree with all data. In particular, the recently released
large $p_T$ data (up to 60-70 GeV) at the LHC are in good agreement
with our predictions. Our results imply that the universality of
color-octet matrix elements holds approximately in charmonium
hadroproduction, when one uses fixed order perturbative calculation
to describe data (the data in small $p_T$ region are not included).
Our work may provide a new test for the universality of color-octet
matrix elements, and the color-octet mechanism in general.

\end{abstract}

\maketitle

\section{Introduction}\label{sec:introduction}

Heavy quarkonium is a multiscale system which can probe various
regimes of QCD. Thus, an understanding of heavy quarkonium
production is particularly interesting. To solve the large
discrepancy between CDF data at the Fermilab
Tevatron\cite{Abe:1992ww} of $\psip$ production at high $p_T$ and
theoretical predictions, the color-octet (CO) mechanism
\cite{Braaten:1994vv} was proposed based on nonrelativistic QCD
(NRQCD) factorization\cite{Bodwin:1994jh}. With the CO mechanism,
$Q\bar{Q}$ pairs can be produced at short distances in CO ($\sa$,
$\sb$, $\pj$) states and subsequently evolve into physical quarkonia
by nonperturbative emission of soft gluons. It can be verified that
the partonic differential cross sections at leading-order (LO) in
$\a_s$ behave as $1/p_T^4$ for $\sb$, and $1/p_T^6$ for $\sa$ and
$\pj$, all of which decrease at high $p_T$ much slower than
$1/p_T^8$ of the color-singlet (CS) state. The CO mechanism could
give a natural explanation for the observed $p_T$ distributions and
large production rates of $\psip$ and $\jpsi$
\cite{Kramer:2001hh,Brambilla:2010cs}. However, CO mechanism seems
to encounter difficulties when the polarization of $J/\psi$ is also
taken into consideration \cite{Affolder:2000nn,Chung:2010iq}. To
exploit the underlying physics, lots of efforts have been made,
either by introducing new
channels\cite{Artoisenet:2007xi,He:2009zzb,Fan:2009zq} or by
proposing other mechanisms\cite{Nayak:2005rw,Haberzettl:2007kj}.

It is a significant step to work out the next-to-leading order (NLO)
QCD correction for the CS channel, which enhances the differential
cross section by about 2 orders of magnitude at high
$p_T$\cite{Campbell:2007ws}, and changes the $J/\psi$ polarization
from being transverse at LO into longitudinal at
NLO\cite{Gong:2008sn}. Although the CS NLO cross section still lies
far below the experimental data, it implies that, compared to the
$\alpha_s$ suppression, kinematic enhancement at high $p_T$ is more
important in the current issue. This observation is also supported
by our recent work\cite{Ma:2010vd} for $\chi_c$ production, where we
find the ratio of production rates of
$\s_{\chi_{c2}}/\s_{\chi_{c1}}$ can be dramatically altered by the
NLO contribution due to change of the $p_T$ distribution from
$1/p_T^6$ at LO to $1/p_T^4$ at NLO in the CS P-wave channels. So we
can conclude nothing definite until  all important channels in
$1/p_T$ expansion are presented. It means the CO channels $\sa$
\cite{Gong:2008ft} and $\pj$ should be considered at NLO, while the
CS channel $^3S^{[1]}_1$  at next-to-next-to-leading order (NNLO) in
$\alpha_s$. Among these corrections, the complete NNLO calculation
for the CS channel is currently beyond the state of the art, and
instead, the NNLO$^\star$ method is
proposed\cite{Artoisenet:2008gk,Lansberg:2008gk}. Compared to NLO,
the only potentially not suppressed contribution within NNLO CS
channel is gluon fragmentation, which gives a new scaling behavior
of $1/p_T^4$ for the cross section. But, as studied in
ref.\cite{braaten}, these fragmentation contributions are ignorable,
compared with experimental prompt production data of $\jpsi$, and we
will further argue about this point in Sec. \ref{sec:NNLO}. As a
result, we expect a complete NLO calculation of $\jpsi$ production
is necessary and sufficient to give a reasonable description of the
experiment data.

Currently, while $\jpsi$ production in two-photon collisions at CERN
LEP2\cite{Klasen:2001cu} and photoproduction at DESY
HERA\cite{Artoisenet:2009xh,Chang:2009uj,Butenschoen:2009zy} are
shown to favor the presence of CO contribution, the $\jpsi$
production at $B$ factories is described well using NLO CS model and
leaves little room for the CO
contributions\cite{Ma:2008gq,Gong:2009kp,Zhang:2009ym,Zhang:2006ay}.
$\jpsi$ production in association with a $W$-boson or $Z^0$-boson at
the LHC is also studied \cite{Gang:2010hc}. However, in all previous
works for heavy quarkonium production, CO long-distance matrix
elements (LDMEs) were extracted at LO, which surfer from large
uncertainties. In order to further test the CO mechanism, it is
necessary to extract CO LDMEs at NLO level. This was studied in our
recent work Ref.\cite{Ma:2010vd} for $\chi_{cJ}$  and
Refs.\cite{talk,Ma:2010yw} for $\jpsi$ and $\psip$. Based on
Ref.\cite{talk}, we further study $\jpsi$ and $\psip$ hadron
production including more detailed discussions in this work.

The remainder of this paper is organized as follows. In Sec.
\ref{sec:fit}, we perform a fit to the CO LDMEs for $\psip$ and
$\jpsi$ using the $p_T$ distributions measured by CDF in
Ref.\cite{Aaltonen:2009dm} and Ref.\cite{Acosta:2004yw}
respectively. In the fit of $\jpsi$, feeddown contributions from
$\chi_{cJ}$ and $\psip$ are considered. We refer interested readers
to Ref.\cite{Ma:2010yw} for details on the calculation and the input
parameters. We will study further theoretical uncertainties in Sec.
\ref{sec:uncertainty}. Then, we compare our predictions with new LHC
data and RHIC data in Sec.\ref{sec:comTh}. After that, a related
work of NLO correction to $\jpsi$ production is compared with ours.
We finally give a brief summary in Sec. \ref{sec:summary}.

\section{Fit Color-octet matrix elements}\label{sec:fit}

\begin{table}
\begin {center}
\begin{tabular}{|c|c|c|c|}
 \hline
 $\sqrt{S} (\tev)$&~~region of $y$~~& ~~~~$r_0$~~~~& ~~~~$r_1$~~~~\\
\hline
1.96 &(~~0~,0.6~) & 3.9 &-0.56\\
\hline
7    &(~~0~,0.75)& 4.0 &-0.55\\
\hline
7    &(0.75,1.50)& 3.9 &-0.56\\
\hline
7    &(1.50,2.25)& 3.9 &-0.59\\
\hline
7    &(~~0~,2.4~) & 4.1 &-0.56\\
\hline
7    &(~~0~,1.2~)& 4.1 &-0.55\\
\hline
7    &(~1.2,1.6~)& 3.9 &-0.57\\
\hline
7    &(~1.6,2.4~)& 3.9 &-0.59\\
\hline
7    &(~2.5,~4~~) & 3.9 &-0.66\\
\hline
7    &(~~2~,2.5~)& 4.0 &-0.61\\
\hline
7    &(~2.5,~3~~)& 4.0 &-0.65\\
\hline
7    &(~~3~,3.5~)& 4.0 &-0.68\\
\hline
7    &(~3.5,~4~~)& 4.0 &-0.74\\
\hline
7    &(~~4~,4.5~)& 4.2 &-0.81\\
\hline
14   &(~~0~,~3~~) & 3.9 &-0.57\\
\hline
0.2   &(~~0~,~0.35~~) & 3.8 &-0.60\\
\hline
0.2   &(~~1.2~,~2.4~~) & 4.0 &-0.66\\
\hline
\end{tabular}
\caption{Experimental conditions with various experimental
collaborations. $r_0$ and $r_1$ are theoretical predictions related
to the short-distance coefficients.}
 \label{tableCondition}
\end {center}
\vspace{-0.5cm}
\end{table}

We find $\pj$ channels have a large K factor and can give important
contributions, thus the $\sb$ channel is no longer the unique source
for the high $p_T$ contribution. In fact, the following
decomposition for the short-distance coefficients holds within an
error of a few percent:
\be \label{decomp} \dspj = r_0~\dssa + r_1~\dssb, \ee
where we find $r_0 = 3.9$ and $r_1=-0.56$ for the experimental
condition with CDF at the Tevatron. $r_{0,1}$ for other conditions
discussed in this work can be found in Table \ref{tableCondition}.
As a result, it is convenient to use two linearly combined LDMEs
\bea \label{MEs} \Majpsi &=\mopa + \dfrac{r_0}{m_c^2}\mopc, \NO\\
\Mbjpsi &=\mopb + \dfrac{r_1}{m_c^2}\mopc, \eea when comparing
theoretical predictions with experimental data for production rates
at the Tevatron and LHC. As pointed out in Ref.~\cite{Ma:2010yw},
although both $\mopb$ and $\dspj$ depend on the renormalization
scheme and the renormalization scale $\muL$, $\Mbjpsi$ is almost
independent of them.

We note that the curvature of experimental cross section is positive
at large $p_T$ but negative at small $p_T$, with a turning point at
$p_T \approx 6\gev$. But the theoretical curvature is always
positive. This implies that data below $7 \gev$ can not be well
explained in fixed order perturbative QCD calculations. If including
these data in the fit, it will cause a large $\chi^2$, which
indicates the fit is not reliable. Therefore, in our fit we
introduce a $\ptcut$ and only use experimental data for the region
$p_T \geq \ptcut$. In the following we use $\ptcut = 7 \gev$.

\begin{figure}
\includegraphics[width=7.5cm]{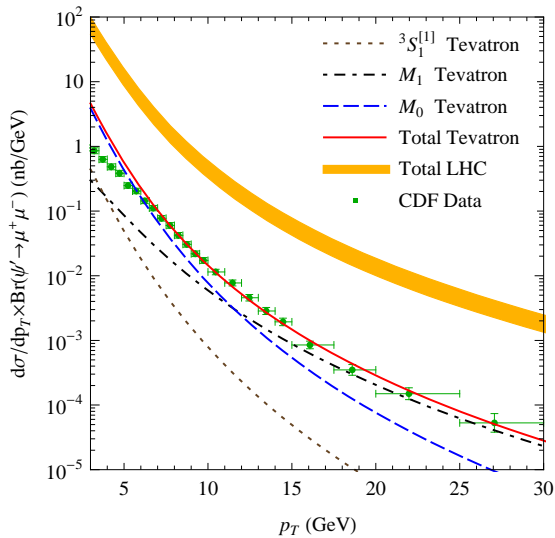}
\caption{\label{fig:psi2s}(Color online.) Transverse momentum distributions of
prompt $\psip$ production at the Tevatron and LHC. CDF data are
taken from Ref.\cite{Aaltonen:2009dm}. The LHC prediction
corresponds to $\sqrt{S} = 14 \tev$ and $|y_\jpsi| <3$.}
\end{figure}

\begin{figure}
\includegraphics[width=7.5cm]{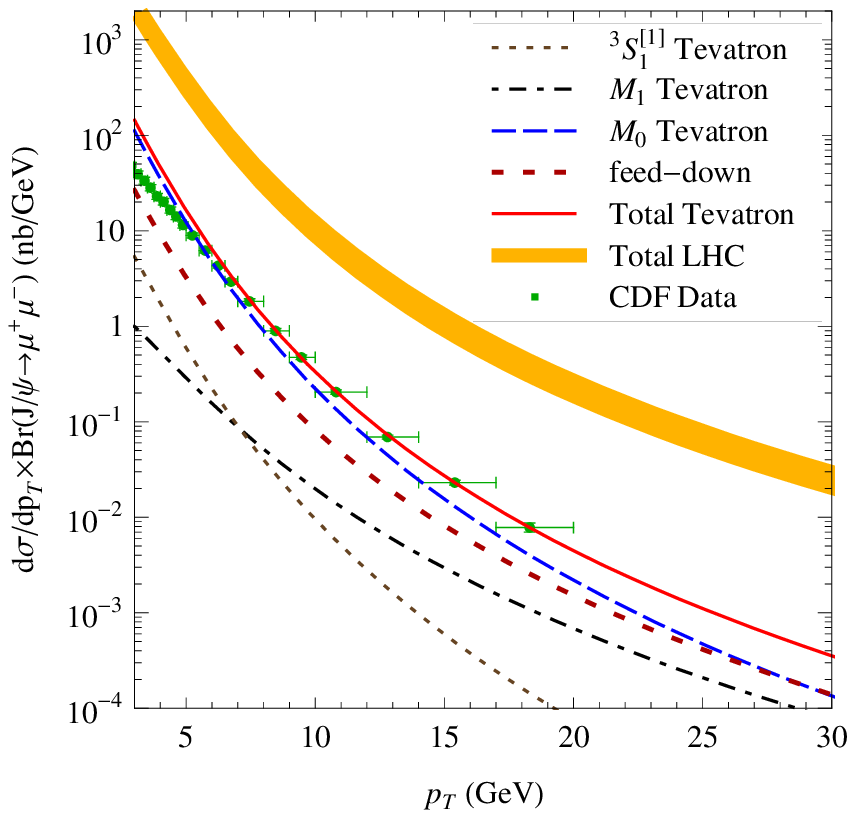}
\caption{\label{fig:jpsi}  (Color online.) Transverse momentum distributions of
prompt $\jpsi$ production at the Tevatron and LHC. CDF data are
taken from Ref.\cite{Acosta:2004yw}. The LHC prediction corresponds
to $\sqrt{S} = 14 \tev$ and $|y_\psip| <3$.}
\end{figure}

\begin{table}
\begin {center}
\begin{tabular}{|c|c|c|c|}
 \hline
$H$&$\MoH$ ($\gev^3$)& $\MbH$ ($10^{-2}\gev^3$)& $\MaH$
($10^{-2}\gev^3$)\\
\hline
$\jpsi$ &$1.16 $&$ 0.05\pm0.02\pm0.02$&$ 7.4\pm1.9\pm0.4$\\
$\psip$&$0.76 $&$ 0.12\pm0.03\pm0.01$&$ 2.0\pm0.6\pm0.2$\\
\hline
\end{tabular}
\caption{Fitted Color-Octet LDMEs in $\jpsi(\psip)$ production with
$\ptcut = 7 \gev$. Here $r_0 = 3.9$, $r_1=-0.56$ are determined from
short-distance coefficient decomposition at Tevatron. The first
errors are due to renormalization and factorization scale
dependence, while the second errors come from the fit. Color-Singlet
($^3S^{[1]}_1$) LDMEs $\MoH$ are estimated using a potential model
result\cite{Eichten:1995ch}.}
 \label{table2}
\end {center}
\vspace{-0.5cm}
\end{table}

By fitting the $p_T$ distributions of prompt $\psip$ and $\jpsi$
production measured at the
Tevatron\cite{Aaltonen:2009dm,Acosta:2004yw} in Fig.~\ref{fig:psi2s}
and Fig.~\ref{fig:jpsi}, the CO LDMEs are determined as showing in
Table \ref{table2}, while the CS LDMEs are estimated using a
potential model result of the wavefunctions at the
origin\cite{Eichten:1995ch}. In Fig.~\ref{fig:psi2s} and
Fig.~\ref{fig:jpsi} we also give the predictions of prompt $\psip$
and $\jpsi$ production at LHC with $\sqrt{S} = 14 \tev$ and $|y| <
3$.

%
%

\section{Theoretic uncertainties}\label{sec:uncertainty}

\subsection{Uncertainty from NNLO color-singlet contribution}\label{sec:NNLO}

Ordinarily, errors come from higher order contributions can be
estimated by varying renormalization scale and factorization scale.
This is the case for CO contributions which have been considered in
the fit. However, for CS contribution, new kinematic enhanced
channels will open at NNLO which behavior as $1/p_T^4$. Because the
new channels have different $p_T$ behavior from LO and NLO
contributions, its influence can not be simply estimated just by
varying parameters at NLO calculation.

A complete NNLO calculation for CS is currently far beyond the state
of the art, instead, a NNLO$^\star$ method is
proposed\cite{Artoisenet:2008gk,Lansberg:2008gk}, in which only tree
level diagrams are considered and an infrared cutoff ($\scut$) is
imposed to control soft and collinear divergences. As $1/p_T^4$
behavior channels are presented for the first time at NNLO, their
contributions do not have divergences and should almost not
dependent on $\scut$ supposing $\scut$ is sufficiently small.
Generally, for small $\scut$ and large $p_T$, the NNLO$^\star$
contributions can be expanded as
 \be \label{eq:nnlo}
{\rm d}\sigma_{\text{NNLO}^\star} = c_4 \frac{1}{p_T^4} + c_6
\frac{\log^2(p_T^2/\scut)}{p_T^6} + ...~~,
 \ee
where ... represents remained contributions which are not important.
To demonstrate terms other than $1/p_T^4$ have negligible
contributions, authors in Ref.~\cite{Artoisenet:2008gk} vary the
$\scut$ and show that the yield ${\rm d}\sigma_{\text{NNLO}^\star}$
becomes insensitive to the value of $\scut$ as $p_T$ increases. The
NNLO$^\star$ contributions are then concluded to be large and
important\cite{Artoisenet:2008gk,Lansberg:2008gk}.

In the following, however, we will argue that the NNLO CS
contribution should not be so large as the NNLO$^\star$ method
expected. We first point out that, there could be a misunderstanding
in Ref.~\cite{Artoisenet:2008gk} when trying to demonstrate the
$1/p_T^4$ term is the most important one. In fact, even if the
second term in Eq.~(\ref{eq:nnlo}) is much larger then the first
term, ${\rm d}\sigma_{\text{NNLO}^\star}$ will also become
insensitive to $\scut$ at large $p_T$, the reason is
 \be \label{eq:appro}
\frac{\log^2(p_T^2/s_{ij}^{'\text{min}})}{\log^2(p_T^2/s_{ij}^{''\text{min}})}
\longrightarrow 1, \text{as $p_T \longrightarrow \infty$ }.
 \ee
Thus it is needed to restudy which term is dominant in ${\rm
d}\sigma_{\text{NNLO}^\star}$ in the current experimental $p_T$
region.

\begin{figure}
\includegraphics[width=9.5cm]{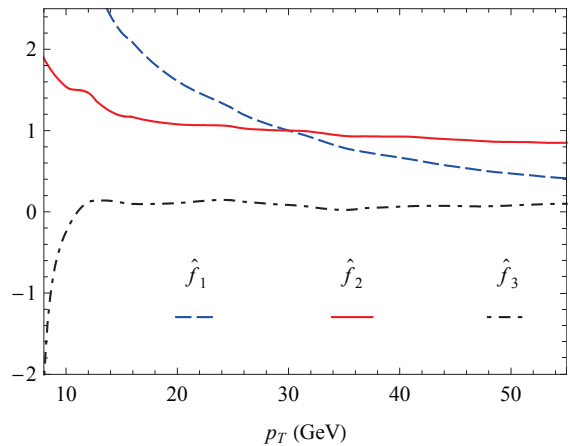}
\caption{\label{fig:NNLO}  (Color online.) Transverse momentum distributions of
functions of $\hat{f}_1$, $\hat{f}_2$ and $\hat{f}_3$. It implies the ${\text{NNLO}}^\star$
result is dominated by the double logarithm enhancement, which will be canceled
in a complete NNLO calculation. See text for definition of $\hat{f}_i$.}
\end{figure}

Our strategy to study this problem is fitting the $p_T$ behavior of
 \bea \label{eq:R}
R^\star = {\rm d}\sigma_{\text{NNLO}^\star} / {\rm d}\sigma_{\text{NLO}},
 \eea
where ${\rm d}\sigma_{\text{NLO}}$ is well known to behave as $1/p_T^6$ at
large $p_T$. If $c_4$ term is dominant in
${\rm d}\sigma_{\text{NNLO}^\star}$, $R^\star$ will behave as $p_T^2$;
while if $c_6$ term is dominant, $R^\star$ will behave as
$\log^2(p_T^2/\scut)$. As there is no difference between $\jpsi$ and
$\Upsilon$ except a mass scale change, we will use the
${\rm d}\sigma_{\text{NNLO}^\star}$ results for $\Upsilon$ in
Ref.~\cite{Artoisenet:2008gk}. Specifically, we define
 \bea \label{eq:R}
f_1&=&\frac{R^\star}{p_T^2}|_{\scut = 0.5 m_b^2},\NO\\
f_2&=&\frac{R^\star}{\log^2(p_T^2/\scut)}|_{\scut = 0.5 m_b^2},
 \eea
while $\hat{f}_1$ and $\hat{f}_2$ correspond to $f_1$ and $f_2$
normalized at $p_T = 30 \gev$. The transverse momentum distributions
of $\hat{f}_1$ and $\hat{f}_2$ are presented in Fig.~\ref{fig:NNLO},
where we find $\hat{f}_2$ is almost fixed to 1 when $p_T > 15 \gev$
while $\hat{f}_1$ still varies significantly in this $p_T$ region.
As a result, the $R^\star$ behaves similar to $\log^2(p_T^2/\scut)$.
To further test this double logarithm behavior, we define
 \bea
\hat{f}_3 = 1 - \frac{R^\star / \log^2(p_T^2/\scut)|_{\scut = 2
m_b^2}}{R^\star / \log^2(p_T^2/\scut)|_{\scut = 0.5 m_b^2}}.
 \eea
It can be found in Fig.~\ref{fig:NNLO} that $\hat{f}_3$ is very
close to 0 when $p_T > 12 \gev$, which confirms our expectation for
the double logarithm behavior.

Based on the above discussion, we may conclude that ${\rm
d}\sigma_{\text{NNLO}^\star}$ in the current experimental $p_T$
region is dominated by $c_6$ term which has double logarithm
enhancement relative to NLO result \footnote{\footnotesize We have
not considered the $\frac{\log^4(p_T^2/\scut)}{p_T^8}$ term in the
expansion in Eq.~\ref{eq:nnlo}, which is important in the region of
$p_T^2 \gtrsim \scut$.}. The double logarithm, originated from IR
cutoff, will be canceled in a complete NNLO calculation with both
real and virtual corrections taken into consideration. Therefore, a
complete NNLO result should have no large enhancement relative to
NLO result\cite{Ma:2010yw}, considering the suppression due to an
extra $\alpha_s$ in NNLO. In other words, the $\text{NNLO}^\star$
method may have overestimated the NNLO contributions.

Having found that the NNLO CS contribution should not be large
relative to the NLO contribution, we may ignore the theoretical
uncertainty from NNLO because the CS NLO result is smaller than
experimental data by at least a factor of 10 at $p_T > 7 \gev$.

\subsection{Uncertainty from decomposing P-wave
channels}\label{uncertanity}

There are two reasons that we should further consider the decomposed
P-wave channel. One is the decomposition in Eq.(\ref{decomp}) is not
exact, although it holds within a few percent, hence we need to
study whether this small error will be enlarged when comparing with
experimental data. The other reason is that $r_0$ and $r_1$ vary
with different center-of-mass energies or different experimental
cuts introduced in experiments, thus the two LDMEs $\MaH$ and $\MbH$
cannot be universally used. Regarding this point, we find the
changes of $r_0$ and $r_1$ are not large in different cases (see
Table \ref{tableCondition}). As a result, $\MaH$ and $\MbH$
extracted from the CDF data can be approximately used to predict
other experimental results. But this can also cause some errors.   A
convenient method to cover all these theoretical uncertainties is
fitting the experimental data using three independent LDMEs. As
pointed out above, data with $p_T < 7 \gev$ may not be well
explained by the fixed order perturbative QCD calculations, so in
the fit we still choose $\ptcut = 7 \gev$, which is safer for the
application of perturbative QCD.

For the $\jpsi$, by minimizing $\chi^2$, we get
 \bea \label{fit3value}
O_1 \equiv \mopa = ~~15.7 \times10^{-2}\gev^3(\pm129\%), \NO \\
O_2 \equiv \mopb = -1.18 \times10^{-2}\gev^3(\pm249\%),  \\
O_3 \equiv \frac{\mopc}{m_c^2} = -2.28
\times10^{-2}\gev^3(\pm239\%).\NO
 \eea
These three LDMEs are unphysically determined, which is reflected by
the large relative errors shown in the end of each expressions.
Nevertheless, it does not matter because we can find some linear
combinations of them, which are physically determined and have small
uncertainties. Define the correlation matrix C
 \bea C_{ij}^{-1} = \frac{1}{2} \frac{{\rm d}^2 \chi^2}{{\rm d} O_i {\rm d} O_j},
 \eea
at the central value points, we have
 \bea
C = \left(
  \begin{array}{ccc}
    0.041 & -0.0060 & -0.011 \\
    -0.0060 & 0.00087 & 0.0016 \\
    -0.011 & 0.0016 & 0.0030 \\
  \end{array}
\right).
 \eea
The eigenvalues $\lambda_i$ with corresponding eigenvectors $\vv_i$
of C are then
 \bea \label{eigen}
 \lambda_1 = 4.5 \times 10^{-2}&,& \vv_1 = (0.96, -0.14, -0.26)\NO \\
 \lambda_2 = 1.2 \times 10^{-6}&,& \vv_2 = (0.29, 0.31, 0.91) \\
 \lambda_3 = 9.2 \times 10^{-9}&,& \vv_3 = (0.047, 0.94, -0.33).\NO
 \eea
The LDMEs corresponding to the eigenvectors are
 \bea \label{DeLambda}
\left(
  \begin{array}{c}
    \EO_1 \\
    \EO_2 \\
    \EO_3 \\
  \end{array}
\right)
 = V \left(
  \begin{array}{c}
    O_1 \\
    O_2 \\
    O_3 \\
  \end{array}
\right),
 \eea
where we denote matrix
 \bea
 V = \left(
  \begin{array}{c}
    \vv_1 \\
    \vv_2 \\
    \vv_3 \\
  \end{array}
\right).
 \eea
Inserting Eqs.(\ref{fit3value}) and (\ref{eigen}) into
Eq.(\ref{DeLambda}), we have
 \bea \label{Lambda}
\EO_1 &=& 15.8 \times10^{-2}\gev^3~~(\pm134\%),\NO \\
\EO_2 &=& 2.11 \times10^{-2}\gev^3~~(\pm5.13\%),  \\
\EO_3 &=& 0.39 \times10^{-2}\gev^3~~(\pm2.45\%). \NO
 \eea
It can be seen that $\EO_2$ and $\EO_3$ are well constrained in this
fit, while $\EO_1$ is badly determined which contains all unphysical
information in Eq.(\ref{fit3value}). Using $\EO_i$, the differential
cross section can be expressed as
 \bea \label{dseq}
{\rm d}\s=\sum\limits_{i=1}^{3}{\rm d}\hat{\s}_i O_i = \sum\limits_{i=1}^{3}a_i
\EO_i, ~~~\text{with}~
\overrightarrow{\textbf{a}}=\overrightarrow{{\rm d}\hat{\s}}~ V^{-1},
 \eea
where ${\rm d}\hat{\s}_i$ denote the corresponding short-distance
coefficients. With its large value and large uncertainty, $\EO_1$
may damage the theoretical results if its coefficient $a_1$ is not
very small. Fortunately, with the CDF condition, we find
contributions of $\EO_1$, $\frac{a_1 \EO_1}{ {\rm d}\s}$, are less than
four percent for all regions of $7 \gev < p_T < 20 \gev$.

In the above treatment, the LDMEs defined in Eq.(\ref{MEs})
correspond to vectors $\vv_{M_0}=(0.25,0,0.97)$ and $\vv_{M_1}
=(0,0.87,-0.48)$, where we have normalized the vectors. We find
$\vv_{M_0} \approx \vv_2$ and $\vv_{M_1} \approx \vv_3$. It means
$\Majpsi$ and $\Mbjpsi$ are approximately equivalent to the two well
constrained ones $\EO_2$ and $\EO_3$ respectively. As a result, if
the badly determined $\EO_1$ is not important, results of using two
LDMEs ($\Majpsi$ and $\Mbjpsi$) and using three LDMEs ($\EO_1$,
$\EO_2$ and $\EO_3$) should be approximately the same.

\begin{figure}
\includegraphics[width=7.5cm]{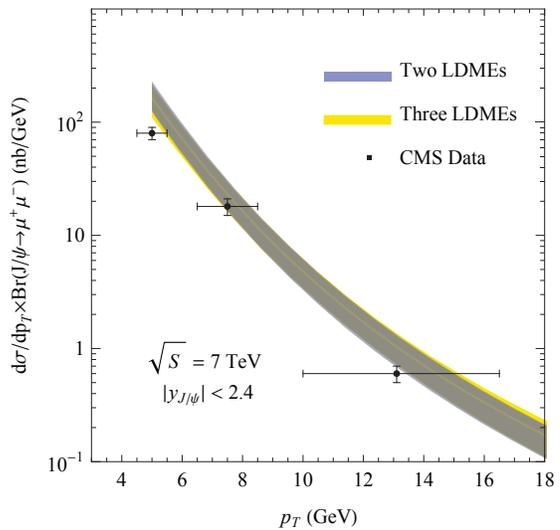}
\caption{\label{fig:cms} (Color online.) Transverse momentum distributions
of prompt $\jpsi$ production at the LHC compared with the CMS data
for $\sqrt{S} = 7 \tev$ and $|y_\jpsi| <2.4$. The CMS data are taken
from Ref.\cite{cms}. The two methods give almost the same
predictions.}
\end{figure}

\begin{figure}
\includegraphics[width=7.5cm]{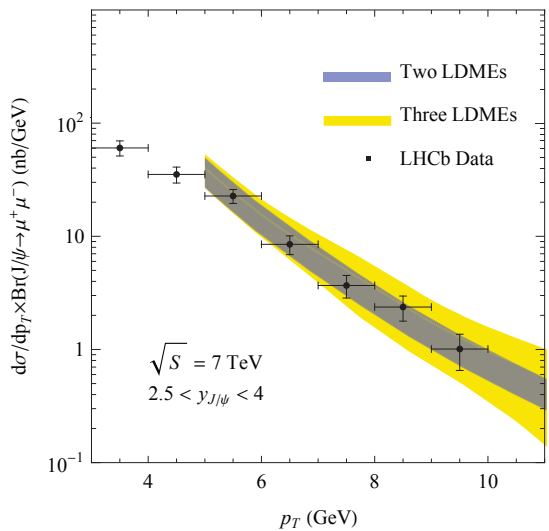}
\caption{\label{fig:lhcb} (Color online.) Transverse momentum
distributions of prompt $\jpsi$ production at the LHC compared with
the LHCb data for $\sqrt{S} = 7 \tev$ and $2.5 < y_\jpsi <4$. The
LHCb data are taken from Ref.\cite{LHCb}. The two methods give the
same predictions for the central values.}
\end{figure}

Comparisons of predictions between using two LDMEs and using three
LDMEs are shown in Fig.\ref{fig:cms} and Fig.\ref{fig:lhcb} for the
measured CMS\cite{cms} and LHCb\cite{LHCb} data respectively.

For the CMS condition ($\sqrt{S} = 7 \tev$ and $\left|y_\jpsi\right|
< 2.4$), we find from Fig.\ref{fig:cms} that the two methods give
almost indistinguishable central values and error bars. This is
understood as $r_{0,1}$ for CMS  only have small differences from
that for CDF, where the LDMEs are extracted. In this case, $a_1$ in
Eq. (\ref{dseq}) is much smaller than $a_2$ and $a_3$, therefore the
contribution of $\EO_1$ is ignorable although it has large
uncertainty. We find that the theoretical predictions are in good
agreement with the CMS data in a very wide range of $p_T$.

For the LHCb condition ($\sqrt{S} = 7 \tev$ and $2.5 < y_\jpsi <
4$), we find from Fig.\ref{fig:lhcb} that although the two methods
give the same central values, the method using three LDMEs have
larger errors when $p_T > 9 \gev$. The reason is the influence of a
relatively large difference of $r_{1}$ between LHCb and CDF (about
$18\%$) on the uncertainty in the method of using three LDMEs  is
enhanced by the large error of $\EO_1$ .
On the other hand, the relatively large difference of $r_{1}$ may
give a chance to extract all three LDMEs with small uncertainties
when experimental data at LHCb are adequate enough. Anyway, it can
be seen from Fig.\ref{fig:lhcb} that our predictions give a good
description for the LHCb data.

In short, the methods of using two LDMEs and using three LDMEs are
consistent in giving predictions in the present situation, when only
two independent LDMEs can be well constrained. The method of using
two LDMEs have advantages of simple formalism and intuitive physical
implication,  as they approximately represent the $p_T^{-6}$ (for
$\Majpsi$) and $p_T^{-4}$ (for $\Mbjpsi$) behaviors of the cross
section, but it needs to consider uncertainties originated from the
decomposition Eq.(\ref{decomp}) and the differences of $r_{0,1}$
additionally. On the other hand, the method of using three LDMEs can
systematically treat all uncertainties but with a more complicated
form, with which it may not be easy to see the physical meaning
directly.

Within the method of using two LDMEs, whether a good prediction can
be achieved is under control from the differences of $r_{0,1}$
between conditions under which we make predictions and conditions on
which the LDMEs are extracted. Because the decomposition in
Eq.(\ref{decomp}) is good  in the cases discussed in this work (see
Table \ref{tableCondition}), we expect there is no large uncertainty
from it.

\section{Predictions for LHC and RHIC}\label{sec:comEx}

\begin{figure*}
\begin{tabular}{ccc}
\includegraphics[width=5.8cm]{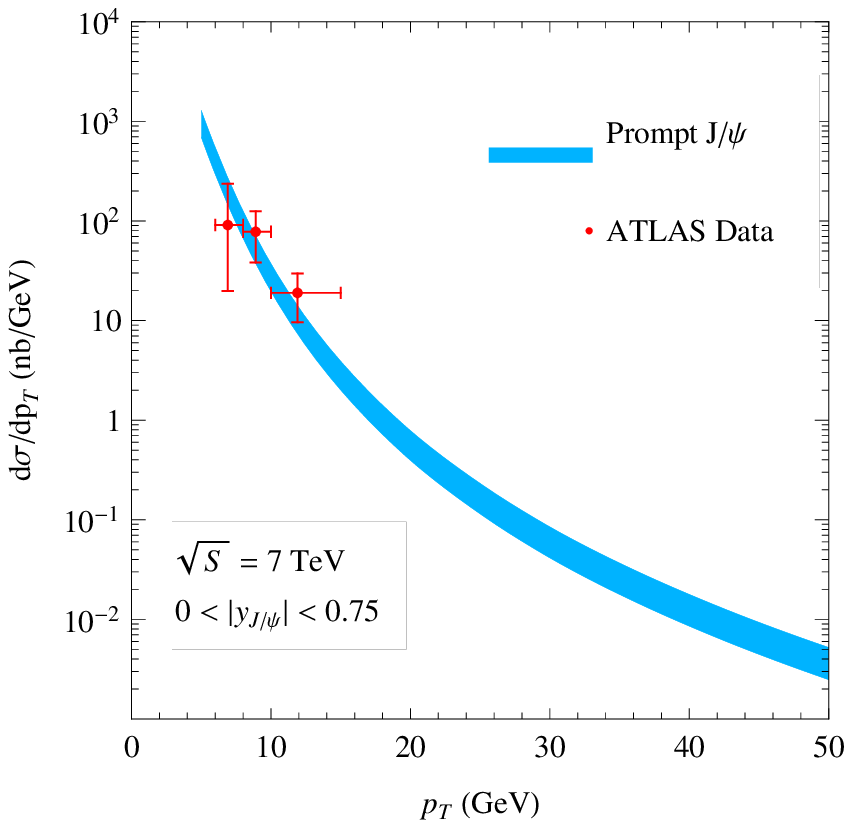}
&
\includegraphics[width=5.8cm]{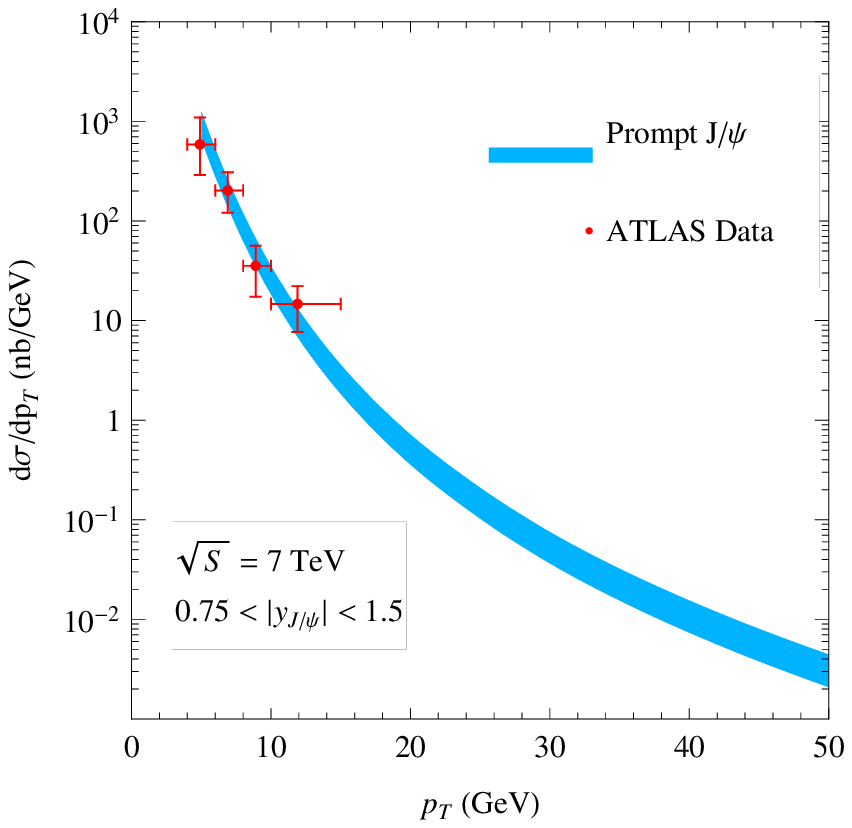}
&
\includegraphics[width=5.8cm]{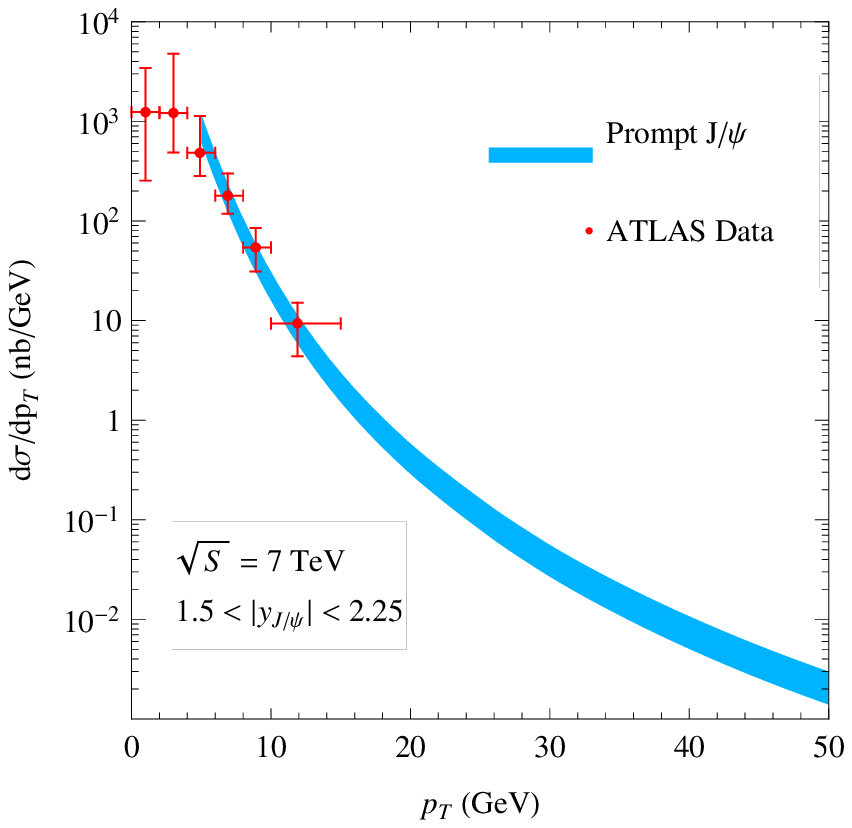}
\\
\includegraphics[width=5.8cm]{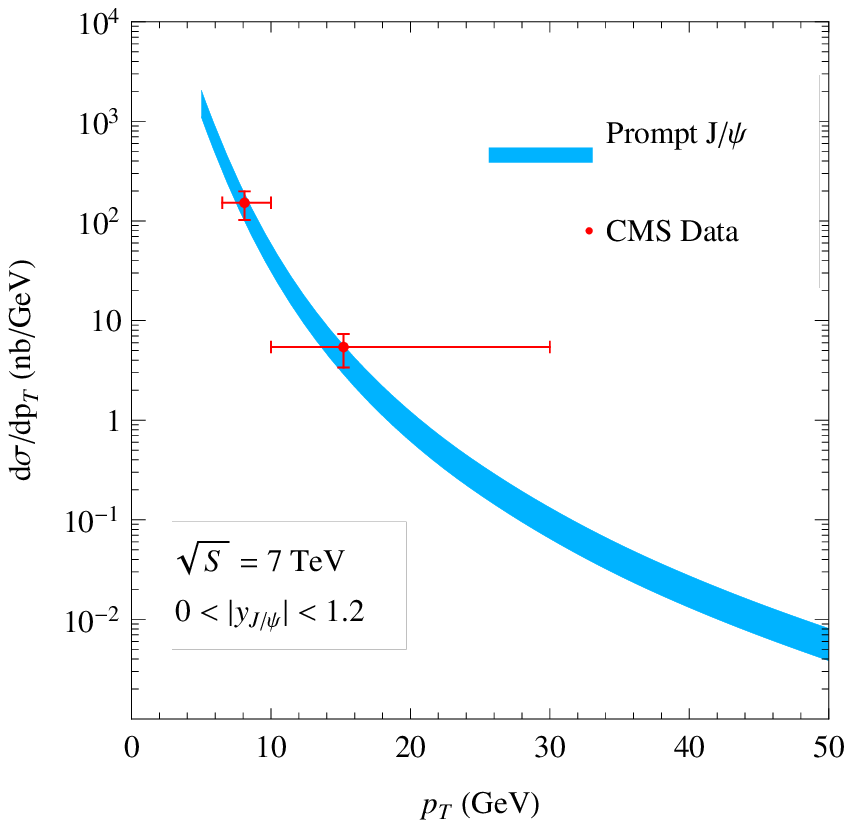}
&
\includegraphics[width=5.8cm]{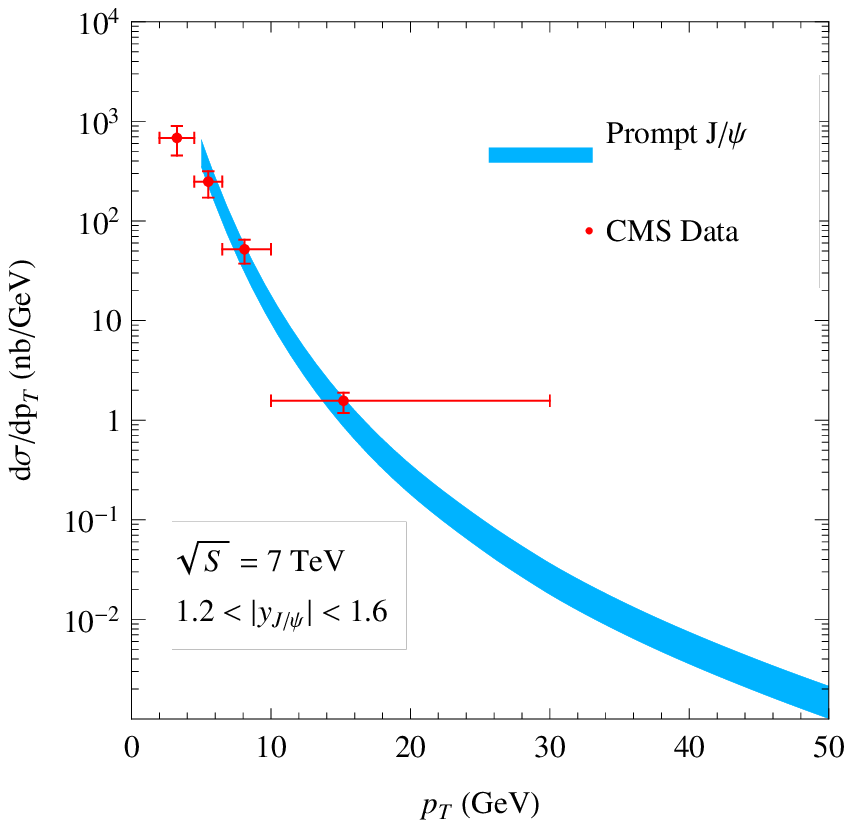}
&
\includegraphics[width=5.8cm]{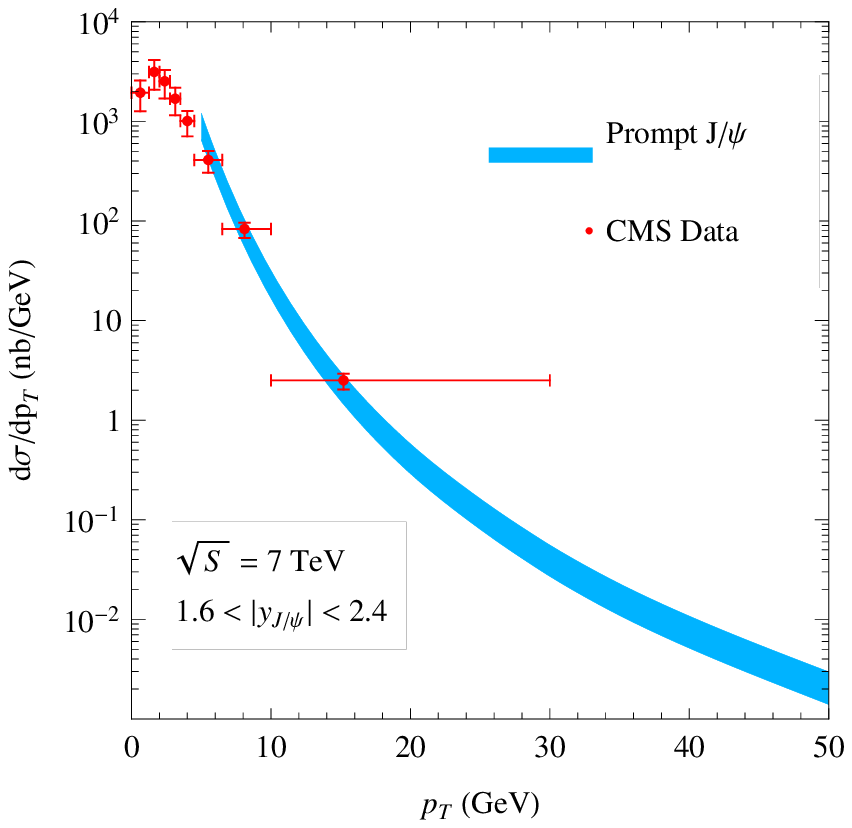}
\\
\includegraphics[width=5.8cm]{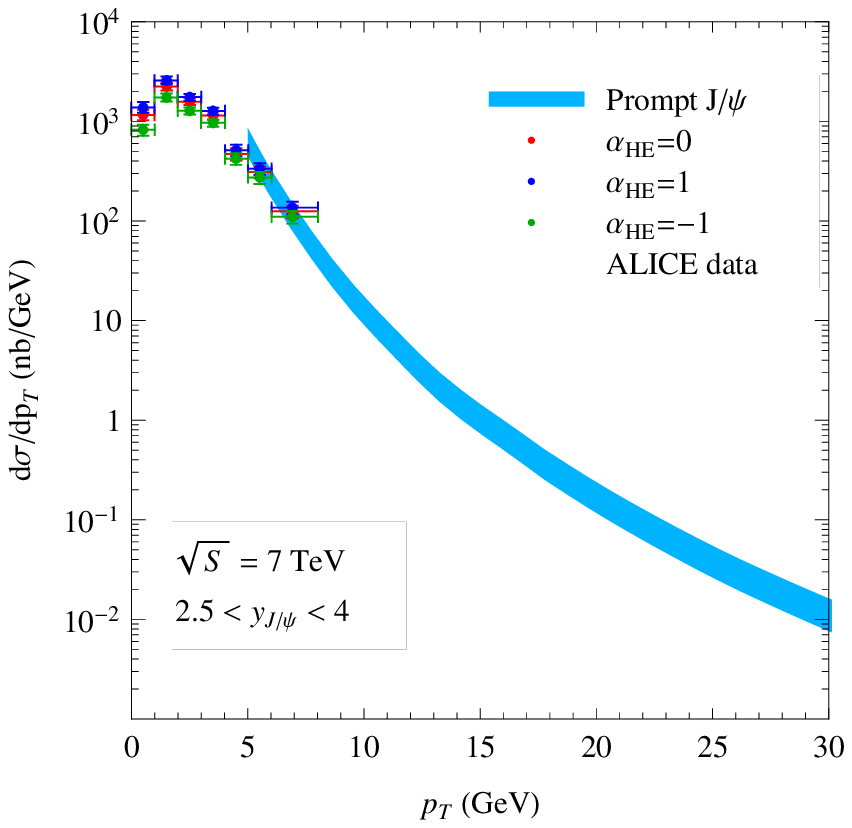}
&
\includegraphics[width=5.8cm]{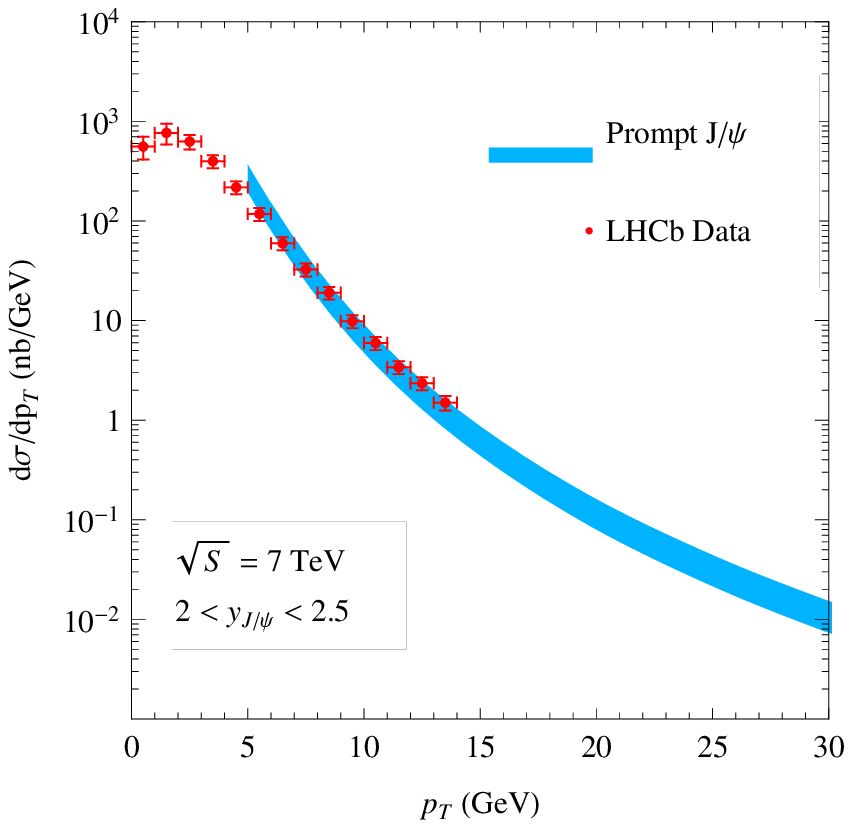}
&
\includegraphics[width=5.8cm]{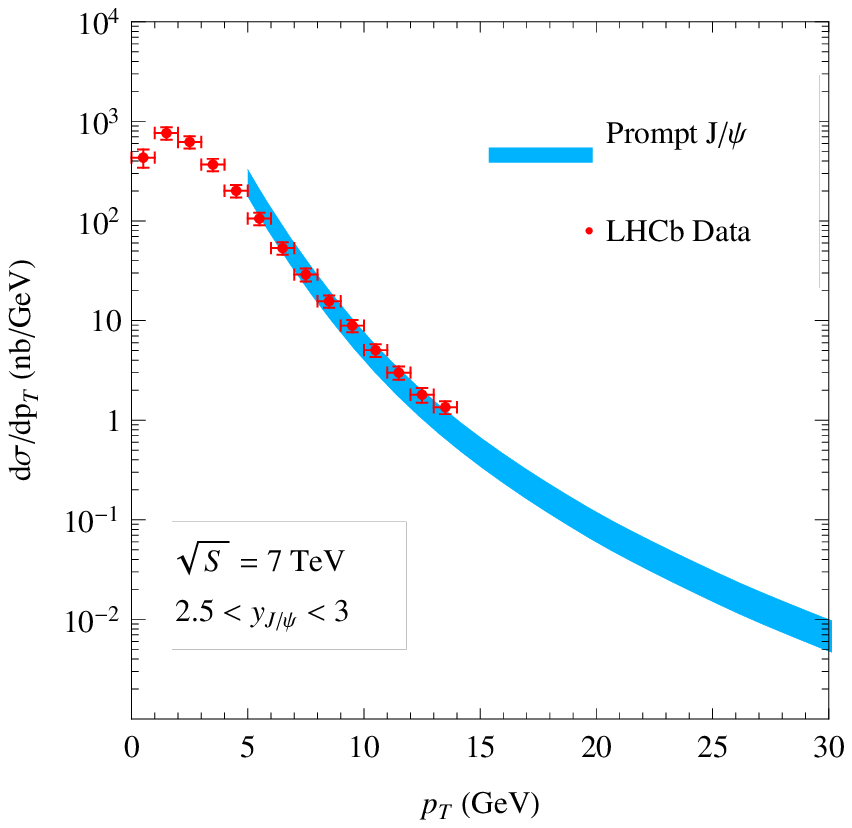}
\\
\includegraphics[width=5.8cm]{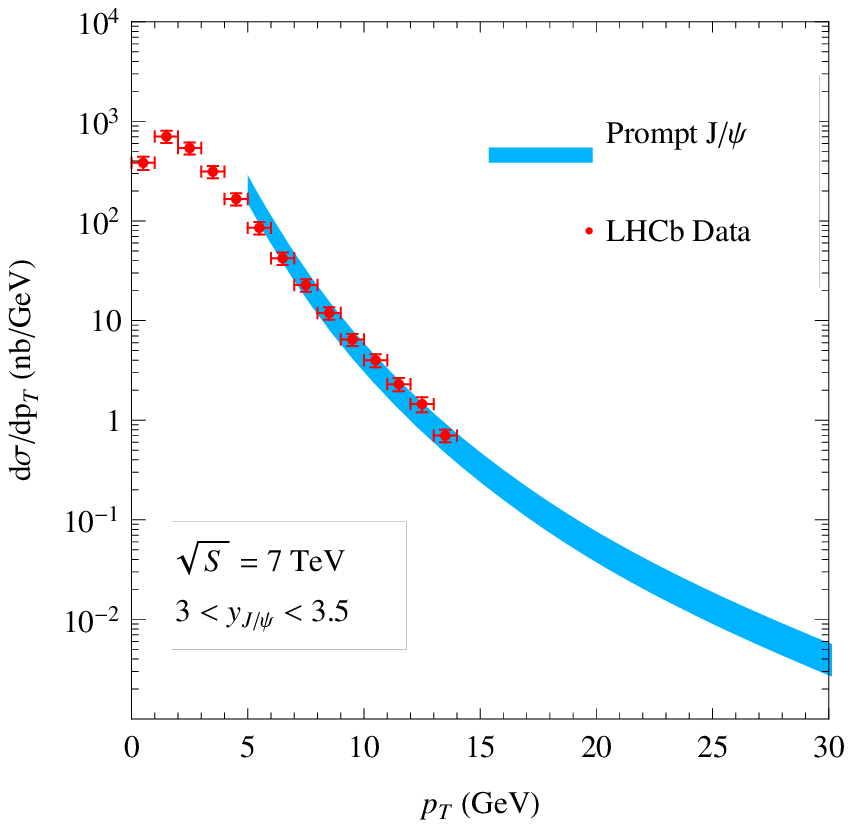}
&
\includegraphics[width=5.8cm]{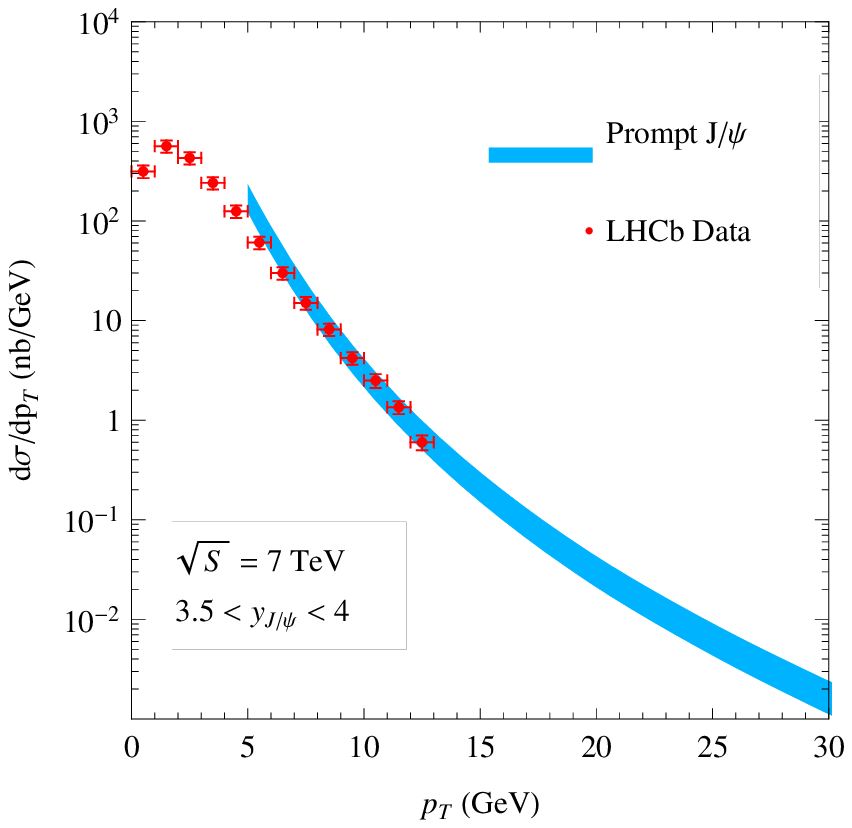}
&
\includegraphics[width=5.8cm]{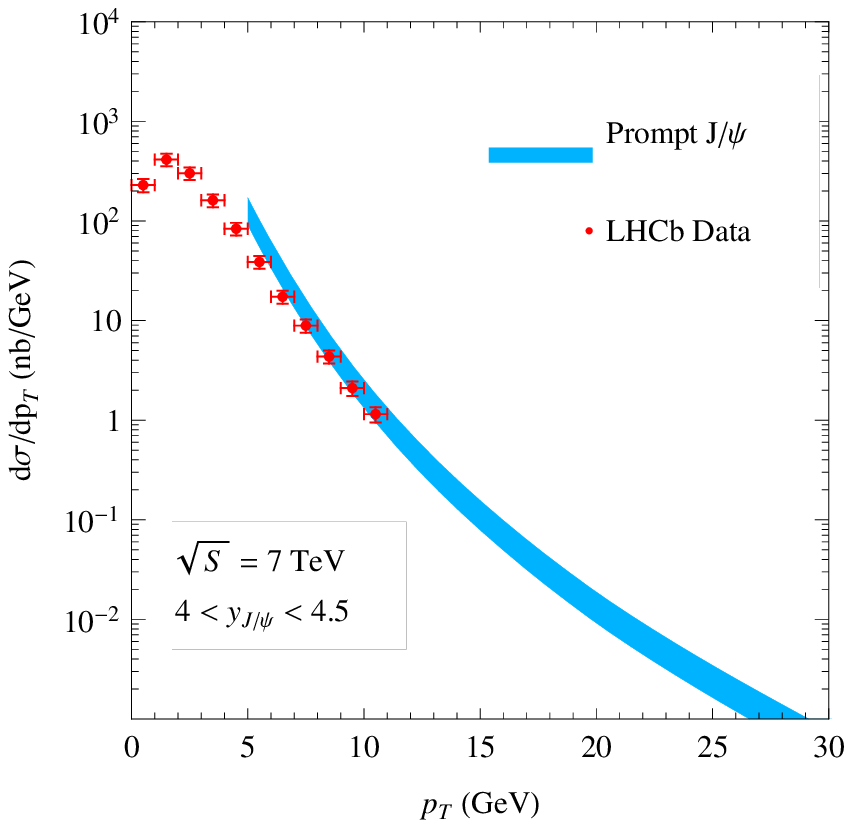}
\end{tabular}
\caption{\label{fig:jpsilhcall} (Color online.) Transverse momentum
distributions of prompt $\jpsi$ production at the LHC compared with
the new data of ALICE, ATLAS, CMS and LHCb Collaborations for
$\sqrt{S} = 7 \tev$. The LHC data are taken from
Ref.\cite{LHC,Khachatryan:2010yr}.}
\end{figure*}

\begin{figure*}
\begin{tabular}{cc}
\includegraphics[width=7.cm]{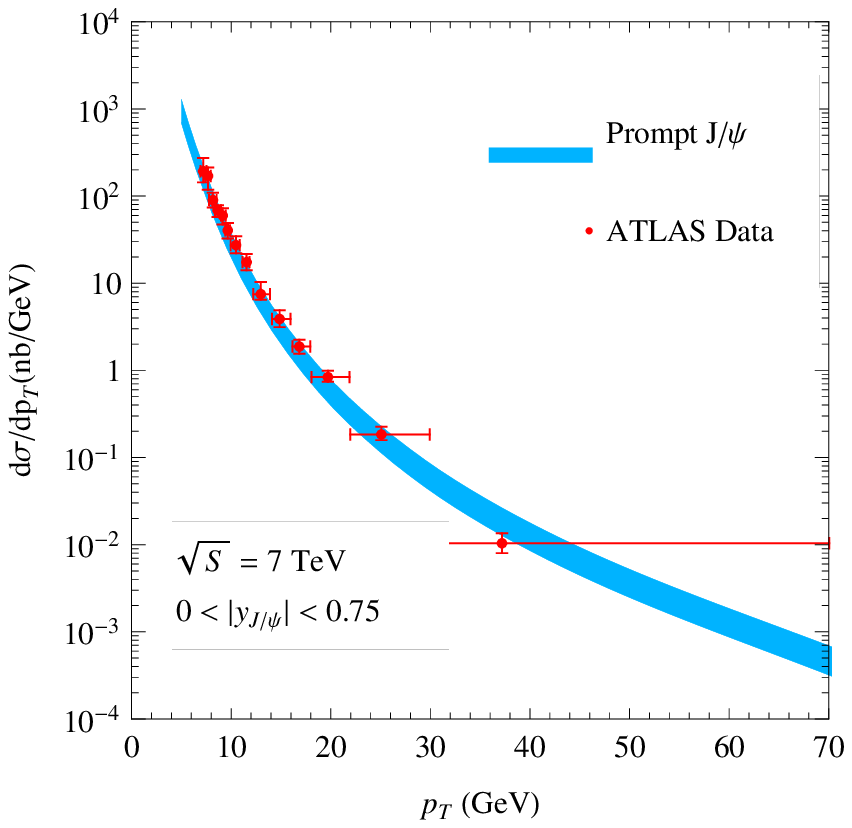}
&
\includegraphics[width=7.cm]{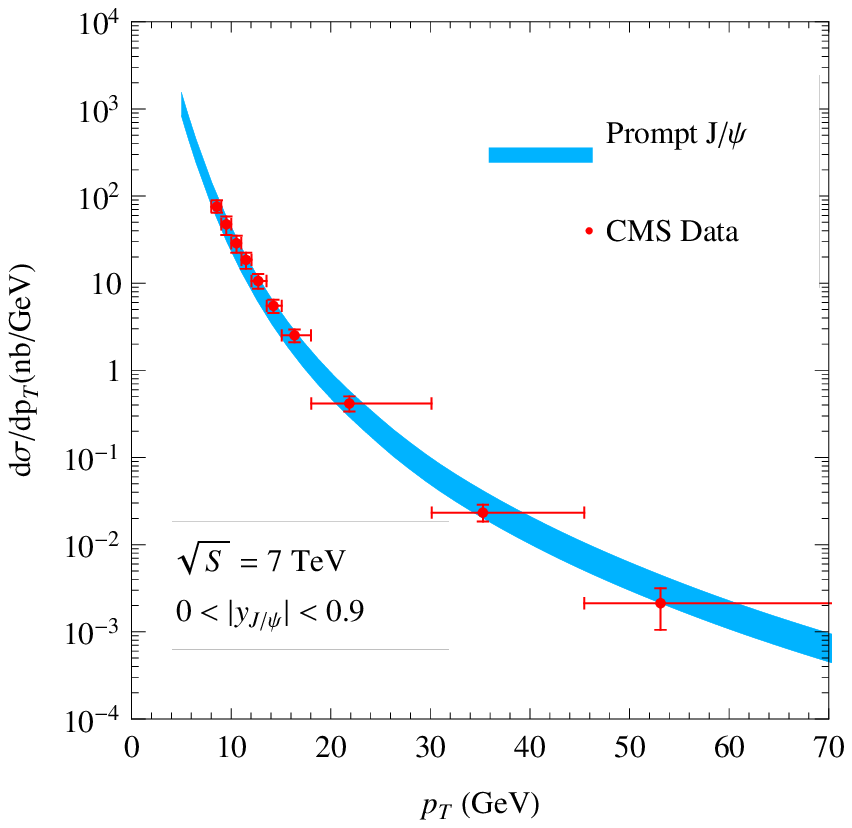}
\end{tabular}
\caption{\label{fig:jpsiLpt} (Color online.) Transverse momentum
distributions of prompt $\jpsi$ production at the LHC in large $p_T$ region.
The ATLAS data are taken from Ref.\cite{ATLASpsi}, and CMS data are taken from
Ref.\cite{CMSpsi}}
\end{figure*}

\begin{figure*}
\begin{tabular}{cc}
\includegraphics[width=7.cm]{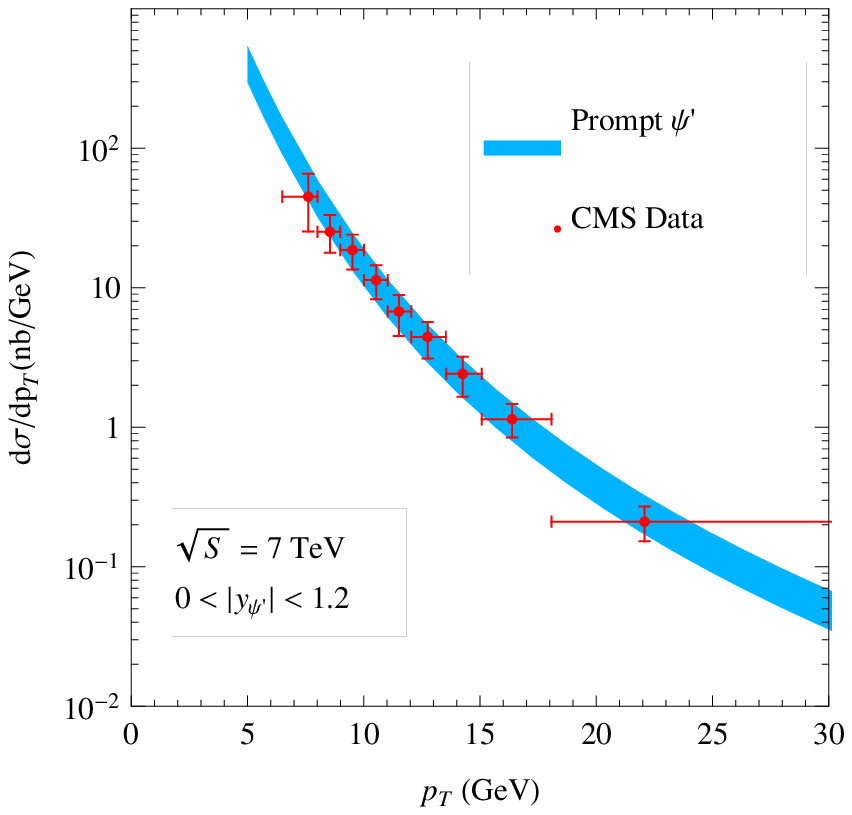}
&
\includegraphics[width=7.cm]{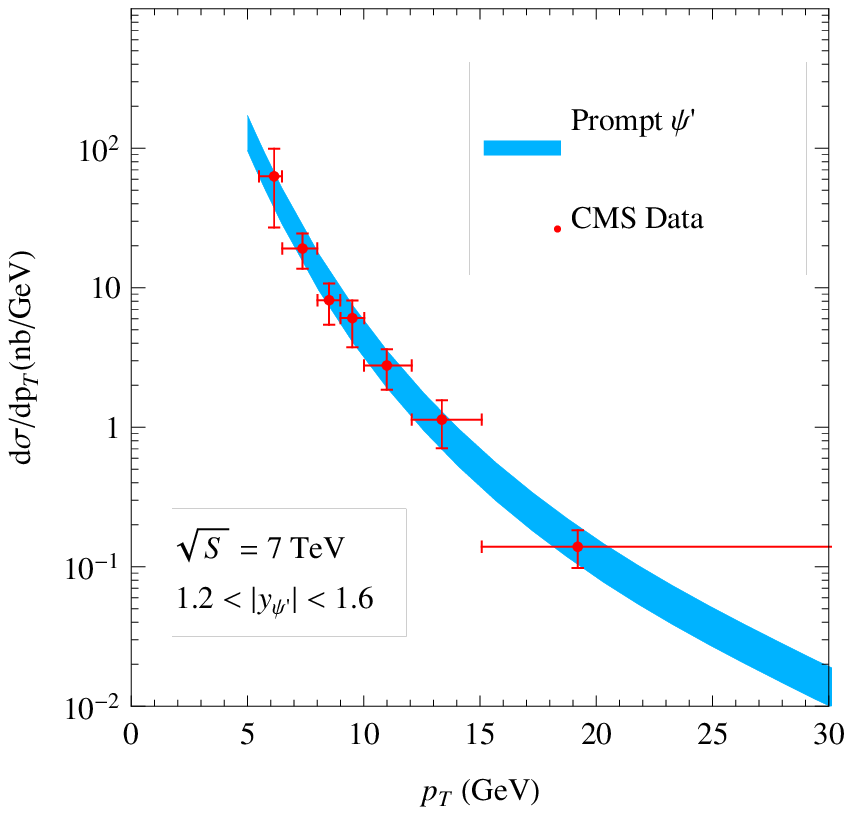}
\\
\includegraphics[width=7.cm]{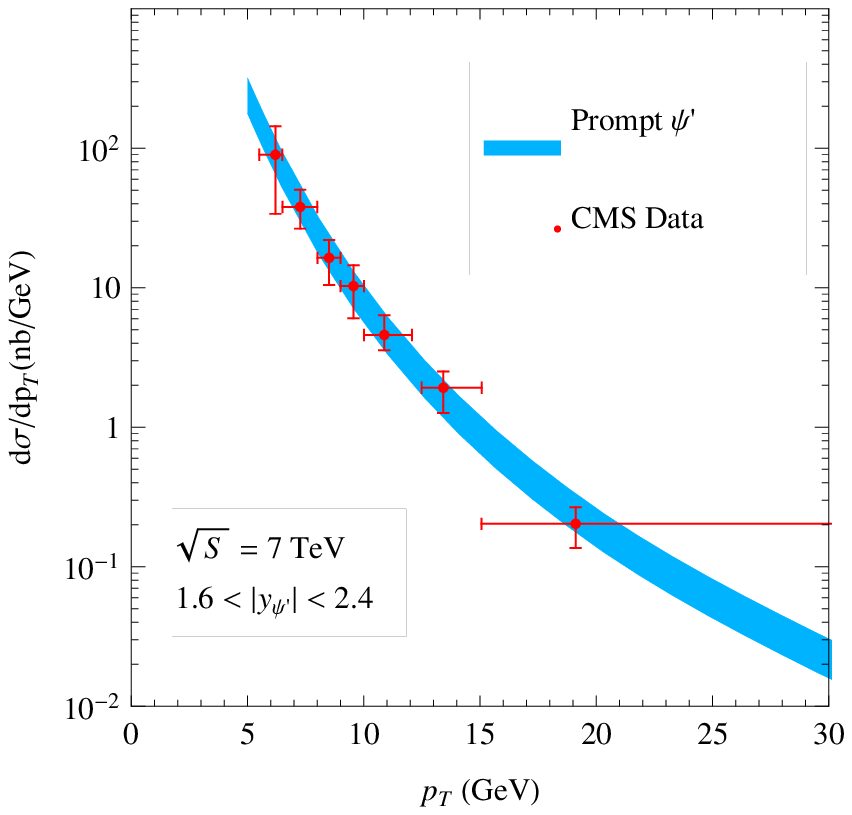}
&
\includegraphics[width=7.cm]{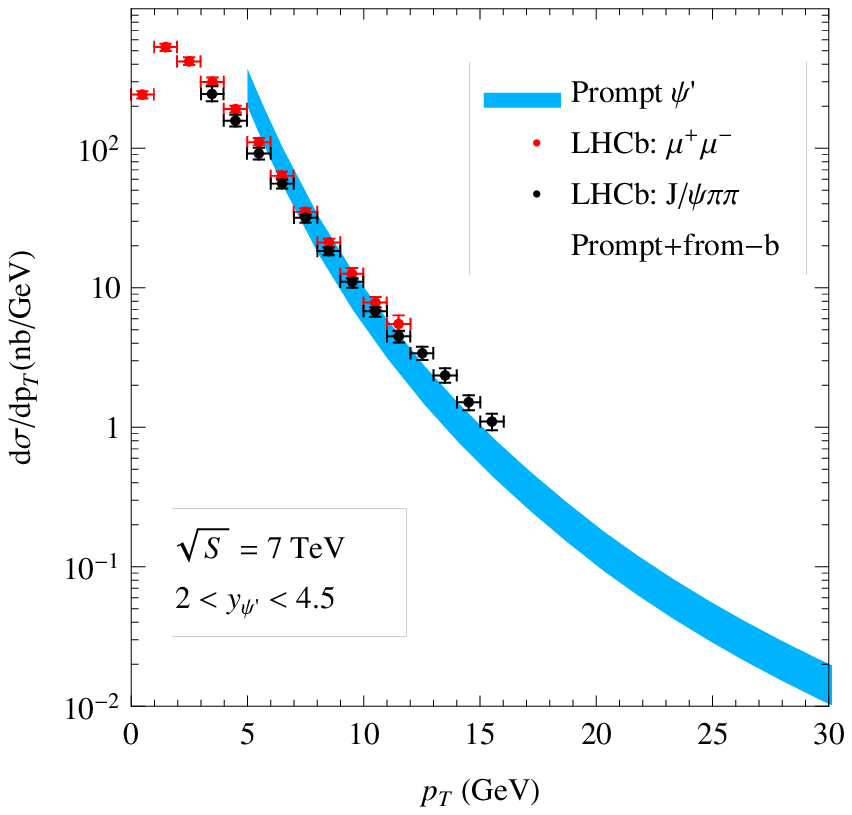}
\end{tabular}
\caption{\label{fig:psi2slhcall} (Color online.) Transverse momentum
distributions of prompt $\psip$ production at the LHC.
The CMS data are taken from Ref.\cite{CMSpsi}, and LHCb data are taken from
Ref.\cite{LHCbpsip}. The LHCb data include also B decay contribution.}
\end{figure*}

\begin{figure*}
\begin{tabular}{ccc}
\includegraphics[width=5.8cm]{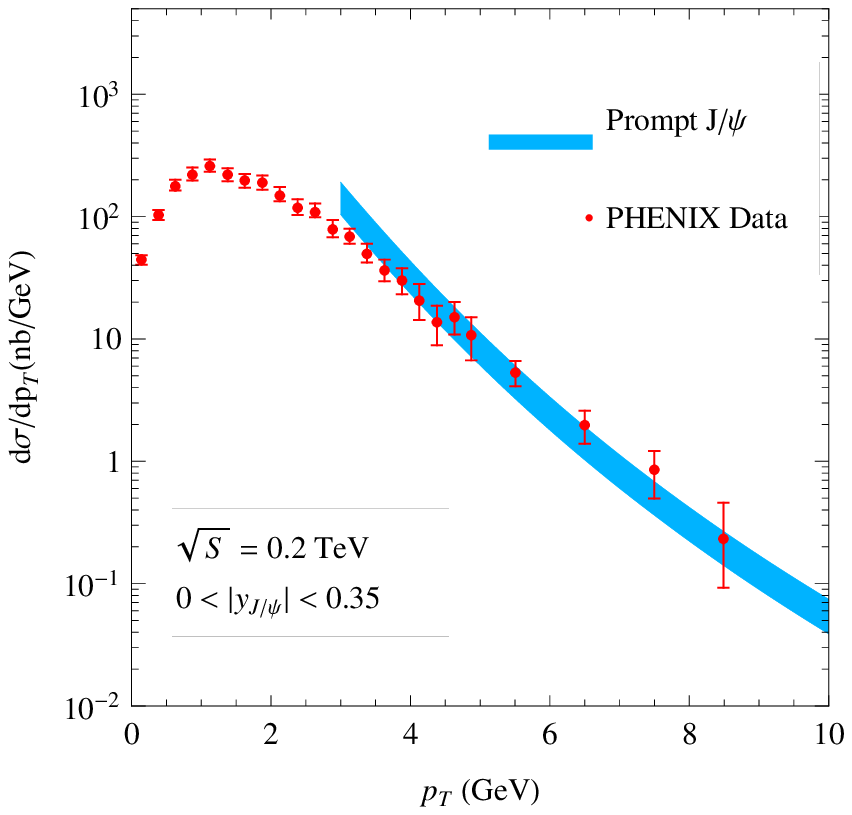}
&
\includegraphics[width=5.8cm]{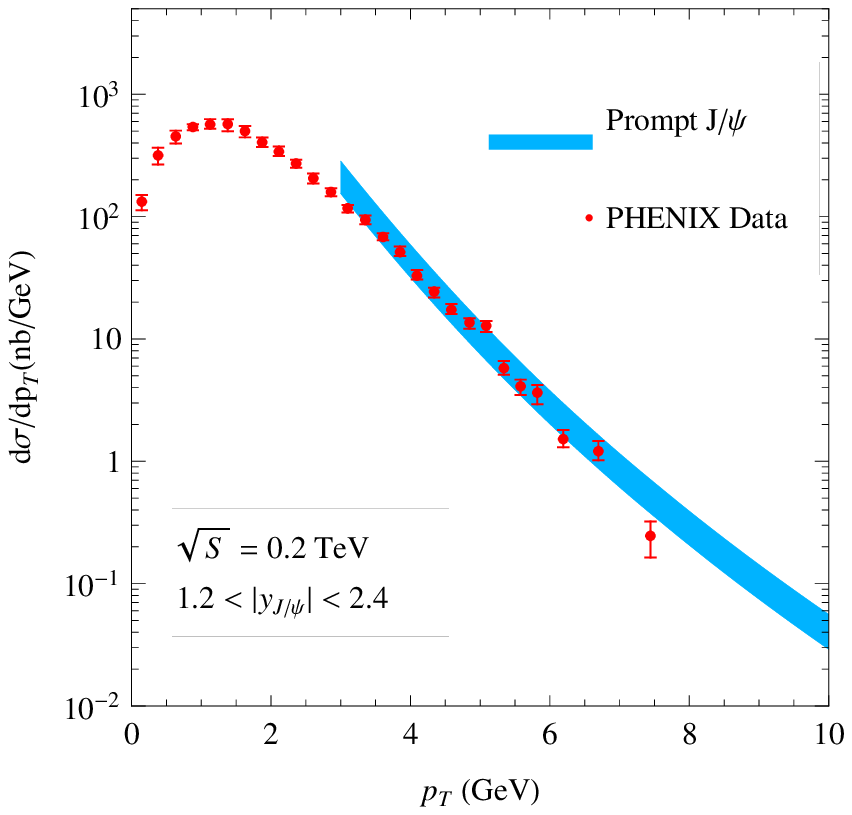}
&
\includegraphics[width=5.8cm]{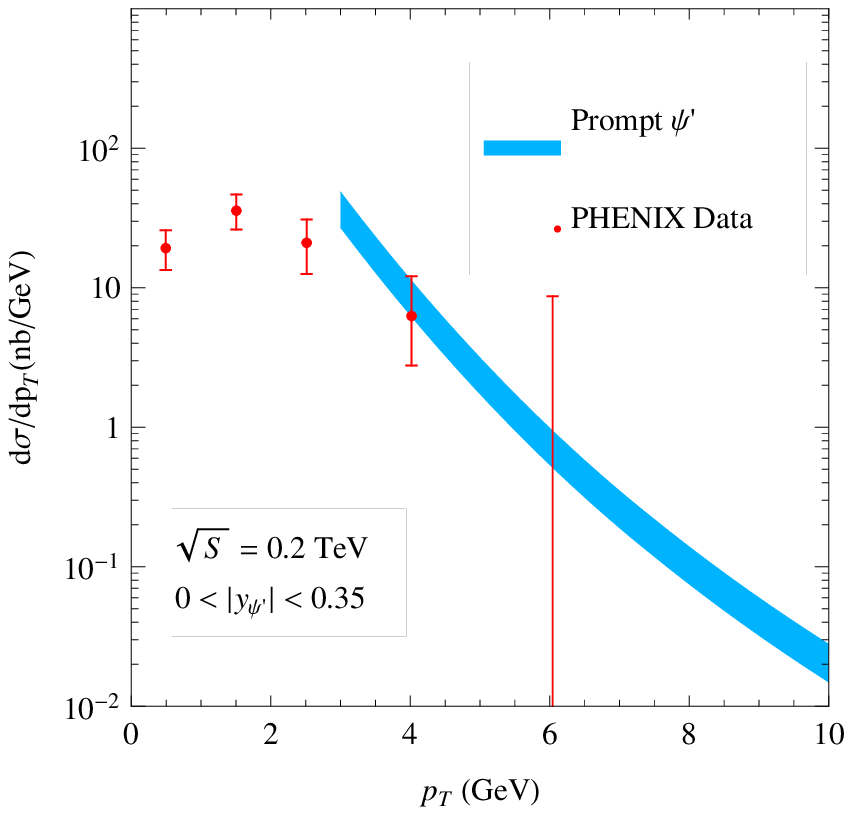}
\end{tabular}
\caption{\label{fig:RHICall} (Color online.) Transverse momentum
distributions of prompt $\jpsi$ and $\psip$ production at RHIC. The
PHENIX data are taken from Ref.\cite{PHENIX}}
\end{figure*}

We compare our predictions of $\jpsi$ prompt production at the LHC
with new LHC data in Fig.\ref{fig:jpsilhcall}. The data of ALICE,
ATLAS and LHCb Collaborations are taken from a recent meeting at
CERN \cite{LHC}, while data of CMS Collaboration are taken from Ref.
\cite{Khachatryan:2010yr}. Besides statistical and systematic
errors, comparable variations from spin-alignment uncertainty are
also considered in data of ALICE, ATLAS and CMS Collaborations.
Errors from spin-alignment are dominant for most $p_T$ points,
therefor, more studies on polarizations are needed in the future. On
the theoretical side, we use the method of using two LDMEs as
discussed in previous sections. It can be found that our predictions
are in good agreement with all data on the whole. Specifically, from
the comparison with the LHCb data, we find predicted cross sections
become declining relative to data as $y_\jpsi$ becomes larger. This
phenomenon, however, can be understood easily because $r_1$ tends to
be far away from $-0.56$ when $y_\jpsi$ becomes larger (see Table
\ref{tableCondition}). On the other hand, as mentioned in the last
section, the relative large difference of $r_1$ may give a chance to
extract all three LDMEs when LHCb has enough data.

Data at large $p_T$ are very important because they may distinguish
between different models. Recently, both ATLAS\cite{ATLASpsi} and
CMS\cite{CMSpsi} Collaborations have released their data of prompt
$\jpsi$ production for $p_T$ as large as $70\gev$. Comparisons with
our predictions (with the same input parameters as in
Refs.\cite{talk,Ma:2010yw}) are shown in Fig.\ref{fig:jpsiLpt},
where it is found that all data are located within predicted
uncertainty bound (a factor of two). We fit the CO LDMEs using the
Tevatron data with $7\gev < p_T <20\gev$ and give a very good
prediction for the LHC data up to $p_T=70\gev$. This is a nontrivial
test for the universality of CO LDMEs. Note that it is certainly
needed to extract the CO LDMEs from these large $p_T$ data when data
are adequate enough.

Our predictions for $\psip$ prompt production at the LHC compared
with CMS data\cite{CMSpsi} and LHCb data\cite{LHCbpsip} are shown in
Fig.\ref{fig:psi2slhcall}. The predictions are in good agreement
with CMS data. For the LHCb, because the data include also $B$ decay
contributions, we can not compare with them directly, but a
consistence between data and prediction can still be found.

We also give predictions for $\jpsi$ and $\psip$ productions at RHIC
in Fig.\ref{fig:RHICall}. It is found that the predictions are in
good agreement with the data.

\section{Comparison with related work}\label{sec:comTh}

Soon after this work was presented in a meeting\cite{talk}, another
talk\cite{Butenschoen:2010px} (see also \cite{Butenschoen:2010rq})
appeared, in which a full NLO QCD correction to direct $\jpsi$
production was also reported. They did not consider feeddown
contributions of $\psi(2S)$ and $\chi_{cJ}$ to $\jpsi$ production,
but jointly fit the Tevatron data and HERA data for $\jpsi$
production (Tevatron data with $\ptcut = 3 \gev$ and HERA data with
$\ptcut = 1 \gev$). It is encouraging that, for all short-distance
coefficients in $\jpsi$ direct production at the Tevatron, results
in our two groups consistent with each other.

However, the results of extracted LDMEs are significantly different.
Specifically, they get \cite{Butenschoen:2010px}
 \bea
\mopa &=& (4.76\pm0.71)\times10^{-2} \gev^3, \NO \\
\mopb &=& (0.265\pm0.091)\times10^{-2} \gev^3,  \\
\mopc &=& (-1.32\pm0.35)\times10^{-2} \gev^5.\NO
 \eea
Inserting them into Eq.(\ref{decomp}), we get
 \bea \label{MabK}
\Majpsi &=& 2.47\times10^{-2} \gev^3, \NO \\
\Mbjpsi &=& 0.594\times10^{-2} \gev^3,
 \eea
which are much different from our results in Table \ref{table2}. The
authors of Ref.\cite{Butenschoen:2010px} also pointed out that
$\Majpsi$ and $\Mbjpsi$ are not precisely corresponding to the well
constrained eigenvectors $\vv_2$ and $\vv_3$ in Eq.(\ref{eigen}),
but also mixed with $\vv_1$, thus in our fit there are very large
uncertainties in LDMEs.

First of all, we note that a small mixing with $\vv_1$ is not so
terrible. If we can expect that the physical LDME corresponding to
$\vv_1$ is not much larger than that corresponding to $\vv_2$ and
$\vv_3$, then the error caused by the mixing is just as large as the
size of mixing, a few percents in our case. When the decomposition
of Eq.~\ref{decomp} holds very well, there will be a LDME which can
only be badly constrained. The fitted value of a badly constrained
LDME is always much larger than its real value because of stochastic
effect, which explains the fact that LDME corresponding to $\vv_1$
is much larger than that corresponding to $\vv_2$ and $\vv_3$ in
Eq.~\ref{Lambda}.

To clarify the discrepancy between Eq.~\ref{MabK} and Table
\ref{table2}, we do a similar fit as authors in Ref.
\cite{Butenschoen:2010px} did: using three LDMEs to fit the Tevatron
data with $\ptcut = 7 \gev$ without considering feeddown
contributions. We then get
 \bea \label{Mab3}
\Majpsi &=& 8.54\times10^{-2} \gev^3~~(\pm12\%), \NO \\
\Mbjpsi &=& 0.167\times10^{-2} \gev^3~~(\pm63\%).
 \eea
Comparing this result with that using two LDMEs to do the fit
without considering feeddown contributions
 \bea \label{Mab2}
\Majpsi &=& 8.92\times10^{-2} \gev^3~~(\pm4.4\%), \NO \\
\Mbjpsi &=& 0.126\times10^{-2} \gev^3~~(\pm18\%),
 \eea
we find the two methods give very similar $\Majpsi$ and $\Mbjpsi$.
Comparing Eq.(\ref{Mab2}) with Table \ref{table2}, we find the
feeddown contributions change $\Majpsi$ a little but reduce
$\Mbjpsi$ by a factor of 2.

We conclude that, even without subtracting feeddown contributions,
results of only fitting Tevatron data with $\ptcut = 7 \gev$ in
Eq.(\ref{Mab3}) are still significantly different from that in
Eq.(\ref{MabK}). Specifically, $\Majpsi$ is well constrained in both
Eq.(\ref{MabK}) and Eq.(\ref{Mab3}), but the central value is much
different. The difference, as short-distance coefficients are the
same and the same fit method is used, must be ascribed to different
treatments for experimental data in the fits. In our opinion, data
for $p_T>3\gev$ at the Tevatron and $p_T>1\gev$ at HERA can not be
described consistently by the fixed order perturbative NRQCD. The
inconsistence may imply that the fixed order perturbative
calculation can not describe the data in small $p_T$ region ($3\gev
< p_T < 7\gev$ for Tevatron and $p_T \sim 1\gev$ for HERA).

Besides, it will be interesting to see if the result given in
Refs.\cite{Butenschoen:2010px,Butenschoen:2010rq} can describe the
large $p_T$ $\jpsi$ production cross sections (say $20\gev < p_T <
70\gev$) observed very recently at the LHC, since the large $p_T$
data provide a very important test for the LDMEs.
%

\section{summary}\label{sec:summary}
In summary,  in this work we calculate the $\jpsi$ and $\psip$
prompt production at the Tevatron, LHC, and RHIC at
$\mo(\a_s^4v^4)$, including all CS, CO, and feeddown contributions.
A large K factor of P-wave CO channels at high $p_T$ results in two
linearly combined LDMEs $\Majpsi$ and $\Mbjpsi$, which can be
extracted at NLO from the Tevatron data. We argue that NLO result is
necessary and essential to give a good description for $\jpsi$
production, because the NNLO CS contributions are unlikely to be so
important as to substantially enhance the cross sections at large
$p_T$. We also compare the method of using two LDMEs with that using
three LDMEs, and find these two methods can give consistent
predictions in the present situation. For $r_{0,1}$, which appear in
two combinations of LDMEs and are related to the short-distance
coefficients depending on given experimental conditions (e.g., the
beam energy, the rapidity values,...), when the differences of
$r_{0,1}$ between the experiment in which the LDMEs are extracted,
and the experiment for which the prediction is made, are small, the
two methods give almost the same predictions with only small errors.
Whereas when the differences are large, predictions of both of the
two methods will have large uncertainties. Our theoretical
predictions are in good agreement with the newly measured LHC data
and RHIC data for both $\jpsi$ and $\psip$ prompt production, which
implies that the universality of CO LDMEs may hold approximately in
charmonium hadroproduction. However, if one uses fixed order
perturbative calculation to describe data in the small $p_T$ region,
we find the universality of color-octet matrix elements may be
broken. Our work provides a new test for the universality of
color-octet matrix elements, and the color-octet mechanism in
general.

\begin{acknowledgments}
We thank G. Bodwin for helpful discussions and suggestions
concerning the error analysis of using three LDMEs. Y. Q. Ma would
also like to thank E. Braaten for useful comments when part of this
result was reported at Topical Seminar on Frontier of Particle
Physics 2010: Charm and Charmonium Physics, Beijing, China, August
27-31, 2010. This work was supported by the National Natural Science
Foundation of China (No.10721063) and the Ministry of Science and
Technology of China (No.2009CB825200).
\end{acknowledgments}




\end{document}